\documentclass[aps, prb, reprint, twocolumn, superscriptaddress]{revtex4-1}
\usepackage{enumitem}
\usepackage{multirow}
\usepackage{graphicx}
\usepackage{longtable}
\usepackage[utf8]{inputenc}
\usepackage[T1]{fontenc}
\usepackage{epstopdf}
\usepackage{textcomp}
\usepackage[cmyk,dvipsnames]{xcolor} 
\usepackage{amsbsy}
\usepackage{amsmath}
\usepackage{amssymb}
\usepackage{amsfonts}
\usepackage{dsfont}
\usepackage{mathtools}
\usepackage{braket}
\usepackage{array}
\usepackage{gensymb}
\usepackage{placeins}
\usepackage{dashrule}
\usepackage{xcolor}
\usepackage{booktabs}
\usepackage{hyperref}
\usepackage{float}
\usepackage{braket}
\usepackage{ulem}

\makeatletter
  \def\my@tag@font{\normalsize}
  \def\maketag@@@#1{\hbox{\m@th\normalfont\my@tag@font#1}}
  \let\amsmath@eqref\eqref
  \renewcommand\eqref[1]{{\let\my@tag@font\relax\amsmath@eqref{#1}}}
\makeatother

\allowdisplaybreaks

\begin{document}

\title{Control and tunability of magnetic bubble states in multilayers with strong perpendicular magnetic anisotropy at ambient conditions}

\author{Ruslan Salikhov}
\email{r.salikhov@hzdr.de}
\affiliation{Institute of Ion Beam Physics and Materials Research, Helmholtz-Zentrum Dresden-Rossendorf, Bautzner Landstrasse 400, 01328 Dresden, Germany}

\author{Fabian Samad}
\affiliation{Institute of Ion Beam Physics and Materials Research, Helmholtz-Zentrum Dresden-Rossendorf, Bautzner Landstrasse 400, 01328 Dresden, Germany}
\affiliation{Institute of Physics, Chemnitz University of Technology, Reichenhainer Strasse 70, 09107 Chemnitz, Germany}

\author{Sri Sai Phani Kanth Arekapudi}
\affiliation{Institute of Physics, Chemnitz University of Technology, Reichenhainer Strasse 70, 09107 Chemnitz, Germany}

\author{Jürgen Lindner}
\affiliation{Institute of Ion Beam Physics and Materials Research, Helmholtz-Zentrum Dresden-Rossendorf, Bautzner Landstrasse 400, 01328 Dresden, Germany}

\author{Nikolai~S.~Kiselev}
 \affiliation{Peter Gr\"unberg Institute and Institute for Advanced Simulation, Forschungszentrum J\"ulich and JARA, 52425 J\"ulich, Germany}

\author{Olav~Hellwig}
\email{o.hellwig@hzdr.de}
\affiliation{Institute of Ion Beam Physics and Materials Research, Helmholtz-Zentrum Dresden-Rossendorf, Bautzner Landstrasse 400, 01328 Dresden, Germany}
\affiliation{Institute of Physics, Chemnitz University of Technology, Reichenhainer Strasse 70, 09107 Chemnitz, Germany}

\date{\today}

\begin{abstract}

The reversal of magnetic bubble helicity through topologically trivial transient states provides an additional degree of freedom that promises the development of multidimensional magnetic memories.
A key requirement for this concept is the
stabilization of bubble states at ambient conditions on application-compatible substrates.
In the present work we demonstrate a stabilization routine for remanent bubble states in high perpendicular magnetic anisotropy [(Co(0.44\,nm)/Pt(0.7\,nm)]$_X$, $X = 48$, 100, 150 multilayers on Si/SiO$_2$ substrates by exploring the effect of external magnetic fields ($H_\mathrm{m}$) of different strength and angles ($\theta$) with respect to the film surface normal.
By systematic variation of these two parameters, we demonstrate that remanent bubble density and mean bubble diameter can be carefully tuned and optimized for each sample. Our protocol based on magnetometry only reveals the densest remanent bubble states at $H_\mathrm{m} = 0.87 H_\mathrm{s}$ ($H_\mathrm{s}$ is the magnetic saturation field) and $\theta=60^\circ - 75^\circ$ for all $X$
with a maximum of 3736 domains/100\,$\mu$m$^2$ for the $X = 48$ sample.
The experimental observations are supported by micromagnetic simulations taking into account the nanoscale lateral grain structure of multilayers synthesized by magnetron sputter deposition, and thus helping understand the different density of the bubble states found in these systems.

\end{abstract}

\maketitle
\section{Introduction}

Noncollinear magnetic textures, such as magnetic domain walls, vortices, bubbles, and skyrmions are currently the subject of intense research. 
New concepts allowing efficient manipulation of these objects at the nanoscale, fuel unrelenting interest in this topic.
Owing to the relatively weak coupling of magnetic spins to the lattice of a host material (thus, avoiding large heat losses), these magnetic textures have the potential for applications in nonvolatile and energy-efficient memory and logic devices, for example, architectures based on artificial neural networks for multidimensional computation~\cite{Sharad,Haung,Yu}. 
In particular, magnetic skyrmions have been suggested for spintronic applications, utilizing the controlled motion of these particle-like magnetic nanotextures~\cite{Muhlbauer,Heinze,Fert,Jiang,Woo,Moreau-Luchaire,YuG,Legrand}.

From the viewpoint of topology, magnetic skyrmions, stabilized by Dzyaloshinskii-Moriya interaction (DMI) are identical to type I magnetic bubbles~\cite{Malozemoff,Milde,Kiselev,Nagaosa}.
The main difference is that skyrmions may have only one energetically favorable chirality, while magnetic bubbles of opposite chirality have identical energies for both chiralities~\cite{Moutafis,Koshibae,Yamane,Koshibae17,Ogawa,Montoya,Montoya2}.
In this respect, magnetic bubbles have an additional degree of freedom to switch between two energetically equivalent states.
The transition between bubbles of opposite chirality typically occurs via a topologically trivial transient state, known as type II bubbles, characterized by an ``onion-like'' magnetic texture and a nonzero net magnetization in the plane of the film hosting the bubble~\cite{Malozemoff,Patek,Hubert}.

The presence of two topological states with opposite chirality and an additional topologically trivial state leads to different dynamical responses of the magnetic bubbles to external driving forces ~\cite{Nagaosa, Moutafis, Koshibae, Yamane, Malozemoff, Koshibae17, Ogawa, Lyu, Montoya}.
Thereby, the magnetic bubbles may provide extra functionality compared to skyrmions stabilized by bulk or interfacial DMI. 
As has been shown earlier, the type I bubbles can be switched to type II bubbles back and forth on a timescale of 100 ps~\cite{Nagaosa, Moutafis, Koshibae}. Moreover, recent studies suggest a new strategy for magnetic memories, employing current-driven bubble helicity reversal~\cite{Hou,Wu}. 

\begin{figure*}[ht]
    \includegraphics[width=1\linewidth]{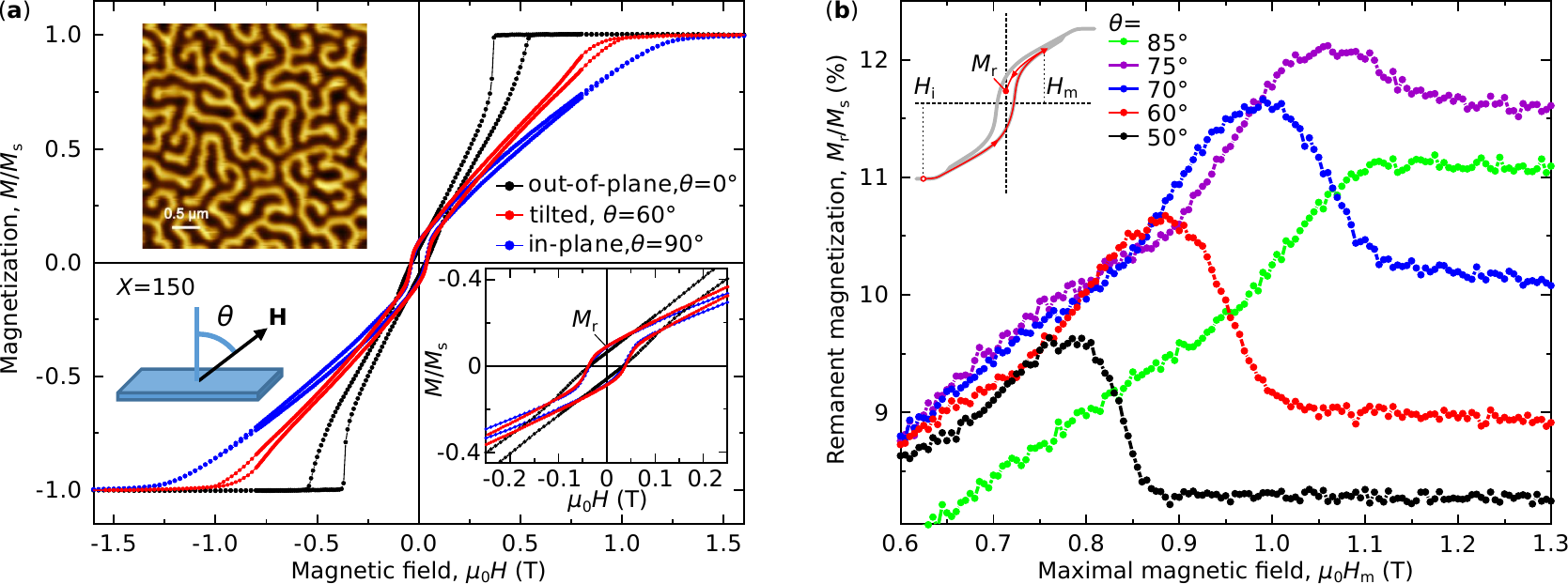}
    \caption{
(a) The magnetic hysteresis loops measured with magnetic field applied parallel (black circles), perpendicular (blue circles) and at tilted angle $\theta=60^\circ$ (red circles) with respect to the [Co (0.44\,nm)/Pt (0.7\,nm)]$_{150}$ multilayer surface normal. The bottom-right inset is an enlarged view of the corresponding hysteresis loops at small magnetic fields. The top-left inset displays a magnetic force microscope (MFM) image of the sample recorded at zero field after OOP saturation. The maze-like stripe domain state is evident in the MFM image. (b) Normalized remanent magnetization ($M_\mathrm{r}/M_\mathrm{s}$) as a function of previously applied magnetic field ($H_\mathrm{m}$) in the DCR protocol for different field angles ($\theta=50^\circ,60^\circ,70^\circ,75^\circ,85^\circ$) with respect to the film normal. In the DCR protocol, the sample was magnetically saturated at negative fields, then a positive field $H_\mathrm{m}$ was applied at an angle $\theta$. In the next step, the field was set back to zero and magnetization was measured at remanence. A characteristic peak of the remanent magnetization at each presented field angle is clearly evident for angles up to $75^\circ$, but vanishes when approaching $90^\circ$, i.e., an in-plane field orientation.      
\label{fig:Fig1}
}
\end{figure*}

This strategy requires steps towards stabilization of dipolar bubbles in metallic multilayers (MLs) at zero field, where the control of a well-defined and strong DMI interaction is not required anymore. 
Such bubble states have been reported in Co/Pt and Fe/Gd MLs with symmetric interfaces~\cite{Chesnel,Fallarino,Desautels,Je}. 
The approach suggested in Refs.~\cite{Montoya2, Chesnel, Fallarino}, however, requires a particular design of the ML to balance the contribution of the perpendicular magnetic anisotropy (PMA), $K_\mathrm{u}$ and the demagnetizing field energy $K_\mathrm{d}=\frac12\mu_0 M_\mathrm{s}^2$ ($\mu_0$ is the vacuum magnetic permeability, and $M_\mathrm{s}$ is the saturation magnetization), which in the case of multidomain states, in addition, scales with the ML thickness.
Noticeably, stable bubble states in all these studies have been obtained in MLs with low quality factor, $Q < 1$, which represents the ratio $Q = K_\mathrm{u}/K_\mathrm{d}$. Furthermore, the magnetic field protocol used for stabilizing bubble states is specific to each material, depending on its magnetic energy characteristics~\cite{Chesnel,Fallarino,Desautels,Je}. 
This brings significant complexity for the stabilization of zero-field magnetic bubbles in metallic MLs.

In this work, we report a reliable approach for the stabilization of magnetic bubbles at zero magnetic field and room temperature in PMA MLs with symmetric interfaces and $Q\ > 1$.
Our approach is based on the search for an optimal magnetic field at different tilt angles ($\theta$) with respect to the ML normal, where the nucleation of bubble state occurs, and can be identified after returning to remanence.
Varying the tilt angle and applied magnetic field
in the same ML, we observe a wide diversity of states with remanent bubbles of different diameters, densities, and arrangements. 
In three samples of significantly different thickness, we find that the optimal range of maximal fields and tilt angles providing the densest bubble states is nearly identical.
Thereby, the presented approach appears suitable for a large family of MLs synthesized by magnetron sputter deposition.

\section{Experiment and discussion}

\begin{figure*}[ht]
    \includegraphics[width=1.0\linewidth]{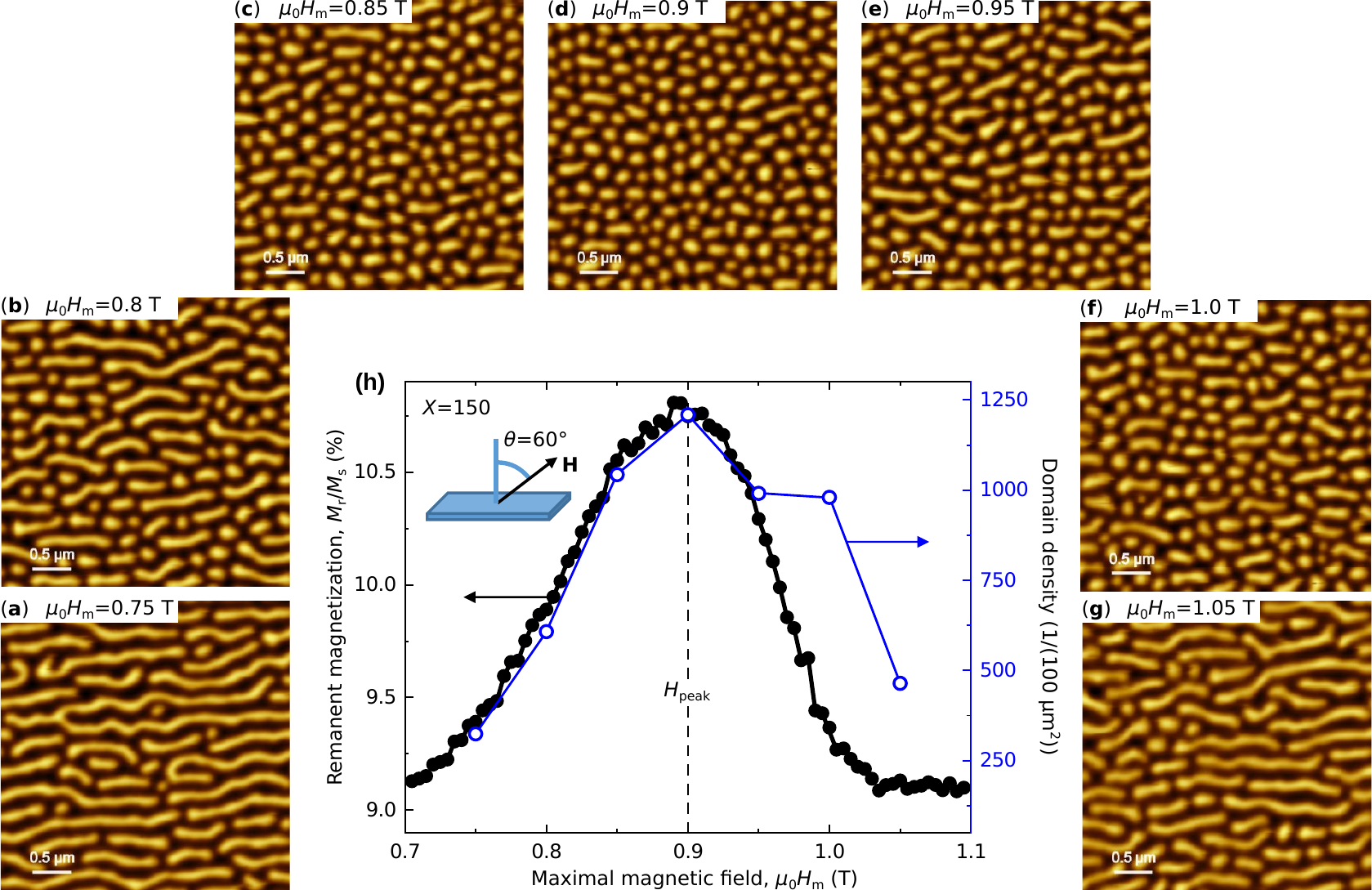}
    \caption{
(a-g) MFM images recorded at zero magnetic field after the application of different magnetic fields $H_\mathrm{m}$ (DCR protocol) at $\theta=60^\circ$ with respect to the [Co (0.44\,nm)/Pt (0.7\,nm)]$_{150}$ film normal. The dense bubble state for $\mu_0H_\mathrm{m}$ = 0.9 T in (d) is evident.
(h) Normalized remanent magnetization (black circles) and magnetic domain density (blue circles)  as a function of previously applied magnetic field ($H_\mathrm{m}$) in the DCR protocol for the field angle $\theta=60^\circ$. Note that the black curve is the enlarged view of the red curve in Figure 1(b). The domain density (blue circles) is calculated from the corresponding MFM images (a-g). It is now visible that the remanent magnetization mimics the domain density. 
\label{fig:Fig2}
}
\end{figure*}

To demonstrate that a dense bubble state can be stabilized in MLs with a high quality factor, $Q > 1$, we select [(Co(0.44\,nm)/Pt(0.7\,nm)]$_X$, $X = 48$, 100, 150 MLs with strong PMA ($Q \approx 1.6$).
Figure~\ref{fig:Fig1}(a) shows representative magnetic hysteresis loops measured in the in-plane (IP) and out-of-plane (OOP) geometries for the sample with $X = 150$. 
The OOP magnetization requires a smaller field for magnetic saturation as compared to the IP loop, which is due to the large PMA.
Both loops exhibit small remanent magnetization at zero magnetic field, $M_\mathrm{r}=M(0)$.
The nonzero $M_\mathrm{r}$ in the OOP geometry results from the imbalance of the volumes of ``up'' and ``down'' domains forming a maze-like pattern shown in the top-left inset in Fig. 1(a)~\cite{Hubert}. 
With increasing the OOP field towards saturation, maze-like domains decay into isolated stripes, which decay further into isolated bubble domains~\cite{Davies}. 
At a critical field, $H_\mathrm{bc}$, slightly below the saturation field, bubble domains collapse. 
When one reduces the field back to zero, the isolated bubble domains experience strip-out instability and expand back into labyrinth domains~\cite{Hubert,Davies}.
The imbalance between ``up'' and ``down'' domains originates from the so-called configuration hysteresis effect~\cite{Hubert} and depends on the magnetic history -- direction and strength of the previously applied fields~\cite{Chesnel,Je}.
On the other hand, the nonzero $M_\mathrm{r}$ in the IP geometry is mainly attributed to the polarization of the Bloch-type domain walls (DWs) in the direction of the external magnetic field~\cite{Salikhov}.
In the case of a tilted magnetic field, $0^\circ<\theta<90^\circ$, both effects, the ratio of ``up'' and ``down'' domains and the domain wall polarization contribute to the remanent magnetization.
It is reasonable to expect that the in-field behavior of the system in the presence of both OOP and IP components of the external field will also differ from the limiting cases of $\theta=0^\circ$ and $\theta=90^\circ$.

As has been shown earlier~\cite{Montoya2, Yu_12, Wu2}, the IP component of the field supports the formation of parallel stripe domains, whereas the OOP component contracts the stripe domains until they pinch off into topologically trivial type-II bubbles.
Such bubbles form a dense irregular lattice shortly before the field reaches the critical value $H_\mathrm{bc}$.
When the bubble domains form a dense enough structure, then, as the external field is decreased back to remanence, labyrinth or stripe domain formation is suppressed due to the high magnetostatic repulsion between the tightly packed bubble domains and the missing space for expansion into stripes~\cite{Charap,Cape}.
Hence, the dense bubble state remains stable even at zero field.
To realize this scenario, one must know i) the optimal tilt angle of the external magnetic field, and ii) the optimal magnetic field strength, at which stripe domains pinch off into bubble domains. 
For this, we employ an approach based on measurements of the remanent magnetization ($M_\mathrm{r}$) as a function of the maximal external field $H_\mathrm{m}$ applied at different tilt angles $\theta$.

\begin{figure*}[ht]
    \includegraphics[width=1.0\linewidth]{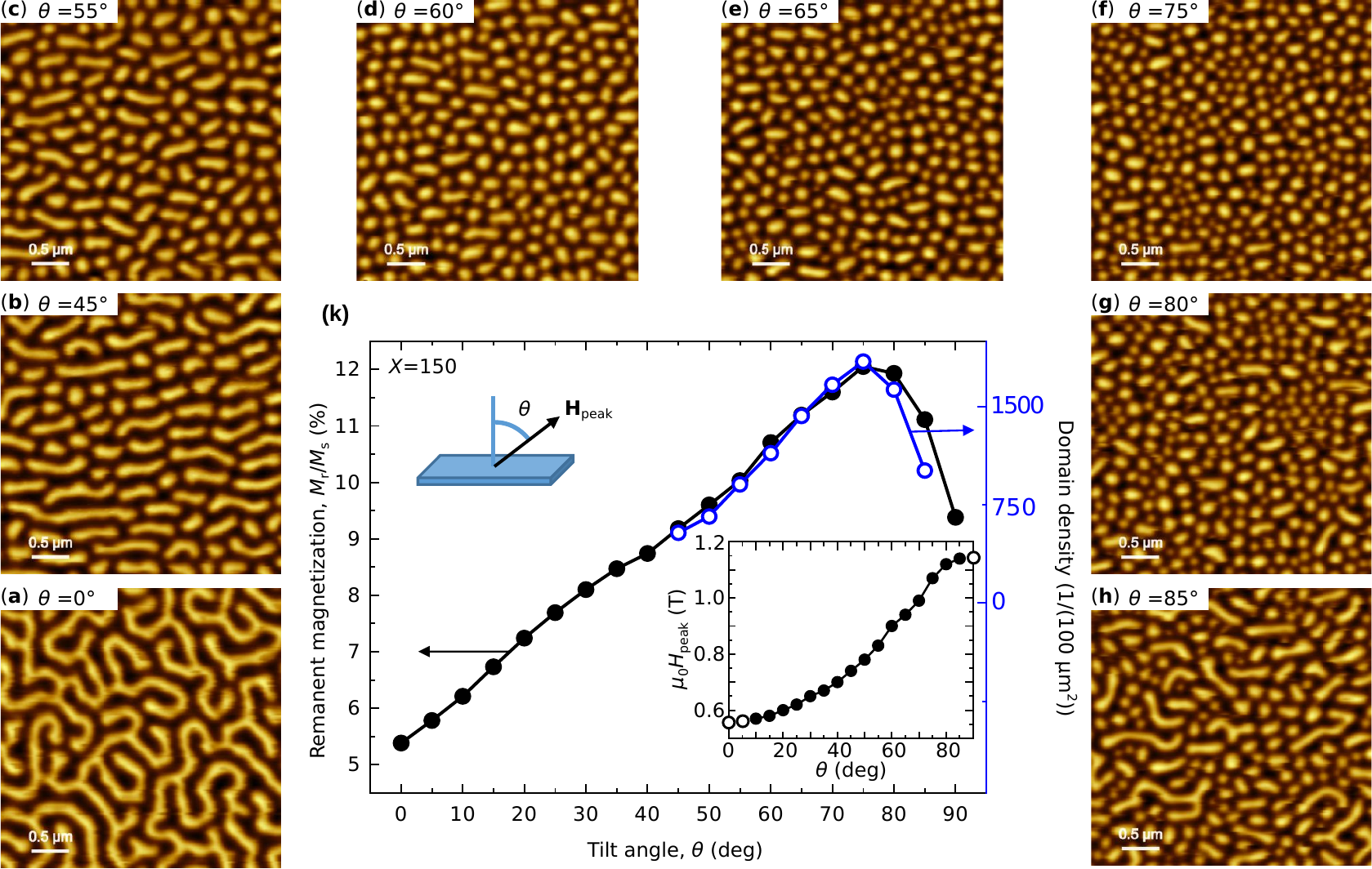}
    \caption{
(a-h) Zero field MFM images, recorded after the DCR protocol for magnetic fields, $H_\mathrm{m}=H_\mathrm{peak}$, which correspond to the maximum remanent magnetization at particular magnetic field angles $\theta$ in the [Co (0.44\,nm)/Pt (0.7\,nm)]$_{150}$ ML.
(k) Maximal remanent magnetization (back circles) and corresponding magnetic domain density (blue open circles) for different $\theta$ in the DCR measurements. The domain density is calculated from the corresponding MFM images shown in (b-h). The inset shows the magnetic fields $H_\mathrm{m}=H_\mathrm{peak}$ as a function of $\theta$.
\label{fig:Fig3}
}
\end{figure*}

Figure~\ref{fig:Fig1}(b) shows the behavior of the normalized remanent magnetization ($M_\mathrm{r} /M_\mathrm{s}$) as a function of $H_\mathrm{m}$ for some selected field angles. 
The measurement protocol (we refer to as DC-remanence (DCR) protocol) is explained by the scheme depicted in the top-left inset in Fig.~\ref{fig:Fig1}(b) and is composed of the following steps.
First, the tilt angle $\theta$ is adjusted and the sample is initially saturated at strong negative field $\mu_0H_\mathrm{i} = -1.8$ T.
Then, a positive field $H_\mathrm{m}$ is applied and after that, the field is gradually (within a few seconds) reduced to zero.
Then $M_\mathrm{r}$ is measured at $H=0$ and plotted as a function of $H_\mathrm{m}$. 
We varied $H_\mathrm{m}$ between 0 T $< \mu_0H_\mathrm{m} < $ 1.6 T with steps of 5 mT. 
In order to maintain identical initial magnetic states, we saturate the sample at negative field $H_\mathrm{i}$ for each successive iteration with the new $H_\mathrm{m}$. 
We note that inverting the sign of the saturation field ($\mu_0H_\mathrm{i}$ = +1.8 T) yields identical behavior.

As follows from Fig.~\ref{fig:Fig1}(b), for any tilt angle with increasing  $H_\mathrm{m}$ the remanent magnetization reaches a plateau.
For $\theta<85^\circ$, the remanent magnetization shows a clear peak shortly before the plateau. 
With increasing $\theta$, the peak and the plateau shift towards higher magnetic fields, because the IP saturation field is larger than the OOP saturation field (Fig.~\ref{fig:Fig1}(a)). 
We obtain the largest value of $M_\mathrm{r}$ at $\theta=75^\circ$, for higher angles, the peak disappears along with an overall reduction of $M_\mathrm{r}$ for the whole range of $H_\mathrm{m}$ (compare, \textit{e.g.}, the curves for $\theta=75^\circ$ and $85^\circ$).

\begin{figure*}[ht]
    \includegraphics[width=1\linewidth]{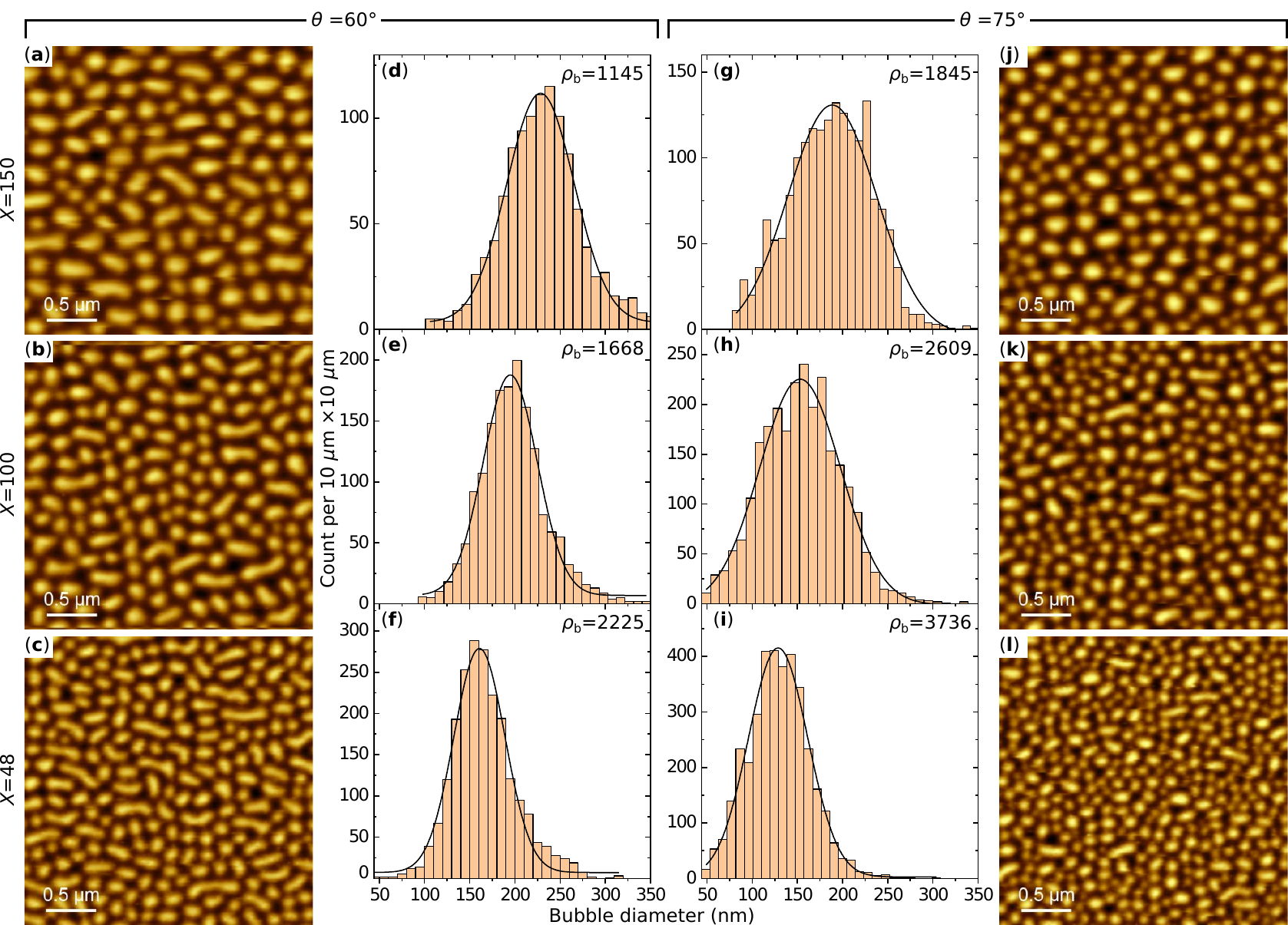}
    \caption{
(a-c) MFM images and  (d-f) size distribution histograms (which include the density of domains $\rho_\mathrm{b}$ per 100 \,$\mu$m$^2$) for magnetic bubbles stabilized at a magnetic field angle of $\theta=60^\circ$ in [Co (0.44\,nm)/Pt (0.7\,nm)]$_X$ MLs with $X =$ 150 (a,d), 100 (b,e) and 48 (c,f). The black solid curve in (d-f) represents the fit to the size distribution using a single Gaussian profile. The Gaussian profile is used for obtaining statistical characteristics, i.e., average bubble diameter $D_\mathrm{b}$ and distribution width $w$ as summarized in Table I.
(g-i) the size distribution histograms for bubbles stabilized at $\theta=75^\circ$, and (j-l) corresponding MFM images for samples with $X =$ 150 (g,j), 100 (h,k) and 48 (i,l).
The solid lines in (g-i) represent a single Gaussian fit, which shows smaller $D_\mathrm{b}$, but broader size distribution as compared to bubbles stabilized at $\theta=60^\circ$.
\label{fig:Fig4}
}
\end{figure*}

In order to verify the domain states across the peak we performed magnetic force microscopy (MFM) measurements for the [Co (0.44\,nm)/Pt (0.7\,nm)]$_{150}$ sample, processed with the DCR protocol, and results are presented in Fig.~\ref{fig:Fig2}. 
Figure~\ref{fig:Fig2}(h) plots the remanent magnetization together with the domain density calculated from the corresponding MFM images shown in Fig.~\ref{fig:Fig2}(a)-(g).
The domain states are stabilized at fixed $\theta=60^\circ$ for different $H_\mathrm{m}$ in the vicinity of the peak. 
It is quite evident that the remanent magnetization can be used as a direct measure for tracking the domain density.
The peak of $M_\mathrm{r}$ in Fig.~\ref{fig:Fig2}(h) corresponds to the densest bubble state, seen in the MFM image at remanence after applying the external field $\mu_0H_\mathrm{m}$ = 0.9 T (Fig.~\ref{fig:Fig2}(d)).
Notably, all MFM images show an alignment of the elongated domains parallel to the IP component of the magnetic field, which indicates the coupling of magnetization within the Bloch type domain walls to the IP component of the external field~\cite{Salikhov}.

To identify the optimal tilt angle, in Fig.~\ref{fig:Fig3}(k) we plot maximal $M_\mathrm{r}/M_\mathrm{s}$ (black circles) and the domain density (blue open circles) as a function of magnetic field angles varying between $\theta=0^\circ$ and $90^\circ$. 
Note that each point in the plot corresponds to $M_\mathrm{r}/M_\mathrm{s}$ at the $H_\mathrm{peak}$ for each individual $\theta$. 
The bottom-right inset in Fig.~\ref{fig:Fig3}(k) shows the dependence of $H_\mathrm{peak}$ on the angle $\theta$.
Since at $\theta<10^\circ$ and $\theta>85^\circ$ the dependence $M_\mathrm{r}(H_\mathrm{m})$ has no peak, in these cases for $H_\mathrm{peak}$ the magnetic field at saturation is chosen, displayed by hollow circles.
Following the angle $\theta$, one sees that the domains in the MFM images Fig.~\ref{fig:Fig3}(a)-(h) evolve from long stripes at $\theta=45^\circ$ to a mixture of short stripes and bubbles at $\theta=55^\circ$ and $60^\circ$.
With further increase of $\theta$ we reach an almost pure bubble state at $\theta=65^\circ$, and with the subsequent increase of the tilt angle up to $\theta=75^\circ$ magnetic bubbles decrease in size leading to an even larger bubble density. 
Thus, by adjusting the field angle, one can control the bubble size and, accordingly, the bubble density. 
The fact that the remanent magnetization mimics the domain density is also evident from the matching of the angular dependencies of the  remanent magnetization and the domain density in Fig.~\ref{fig:Fig3}(k).
The higher the asymmetry between up and down remanent domain areas, the higher is the resulting remanent magnetization.
The highest symmetry and lowest magnetization is reached in a labyrinth stripe domain state where the inversion of up and down domains basically yields identical average states, except for a very small hysteretic difference in domain width~\cite{Davies}.
The highest asymmetry is obtained in a dense bubble state, where lots of domains of one polarity are surrounded by a single interconnected domain of the other polarity. This very frustrated state has the highest remanent magnetization and is stabilized by the magnetostatic repulsion of the bubbles.
Due to this relationship between remanent moment and micromagnetic domain state, the remanent magnetization is a very reliable indicator for the presence of magnetic bubble states.

In order to verify the applicability of our approach for bubble states in different ML systems, we use the same DCR protocol for thinner [Co (0.44\,nm)/Pt (0.7\,nm)]$_X$ samples with $X = 100$ and $48$. 
The corresponding MFM images, bubble densities $\rho_\mathrm{b}$, and size distributions for bubbles, stabilized at $\theta=60^\circ$ and $\theta=75^\circ$, are displayed in Fig.~\ref{fig:Fig4} for all samples.
The equivalent bubble diameter is calculated from the domain area, assuming a circular shape. All distributions show characteristic peaks, which we fit using a single Gaussian profile. The fit parameters are summarized in Table I.
It is apparent that the bubble domain size decreases with decreasing film thickness as expected for MLs with PMA at this thickness regime~\cite{Chesnel,Fallarino,Hellwig}.
The highest bubble density is obtained for the X = 48 sample at an angle of $\theta=75^\circ$ (Fig.~\ref{fig:Fig4}(i) and Fig.~\ref{fig:Fig4}(l)) with $\rho_\mathrm{b}$ = 3736 domains per 100\,$\mu$m$^2$.
All samples reveal that the mean bubble diameter, $D_\mathrm{b}$, for states stabilized at larger field angles ($\theta=75^\circ$) is smaller than for states stabilized at smaller angles ($\theta=60^\circ$). The latter states, however, show broader size distribution, as indicated by the full width at half maximum ($w$) of the Gaussian profile (see Table I).
We obtain the smallest bubble diameter ($D_\mathrm{b}$ = 125 nm) in the $X = 48$ sample, however, the bubble size distribution remains broad ($w$ = 65 nm). A small portion of bubbles with a diameter of about 50 nm can be stabilized using the DCR protocol at $\theta=75^\circ$.
In order to illuminate the physical origin of the broad size distribution of bubble domains in our samples, we perform micromagnetic simulations using the mumax code~\cite{mumax}.

\begin{table}[ht]
\caption{\label{Tab} The total ML film thickness, the mean value of diameter $D_\mathrm{b}$, and size distribution width $w$ for magnetic bubbles stabilized at $\theta=60^\circ$ and $\theta=75^\circ$ field angles in [Co (0.44\,nm)/Pt (0.7\,nm)]$_X$ MLs with $X =$ 48, 100 and 150.}
\begin{tabular}{p{0.07\textwidth}p{0.08\textwidth}p{0.07\textwidth}p{0.08\textwidth}p{0.07\textwidth}p{0.07\textwidth}}
\hline
\hline
Sample      & ML thickness  & $D_\mathrm{b}$,\,nm  & $w$,\,nm  & $D_\mathrm{b}$,\,nm  & $w$,\,nm   \\
\hline
              &               & $\theta=60^\circ$   &     & $\theta=75^\circ$  \\
\hline
X = 48        & 55\,nm      & $160\pm15$     & $55\pm10$    & $125\pm10$    & $65\pm10$ \\
\hline
X = 100       & 114\,nm     & $185\pm15$     & $55\pm10$    & $145\pm15$   & $85\pm10$  \\
\hline
X = 150       & 171\,nm     & $225\pm15$     & $75\pm10$    & $180\pm15$    & $95\pm15$ \\

\hline
\hline
\end{tabular}

 \end{table}

\section{Micromagnetic simulations}

\begin{figure}[ht]
    \includegraphics[width=8.5cm]{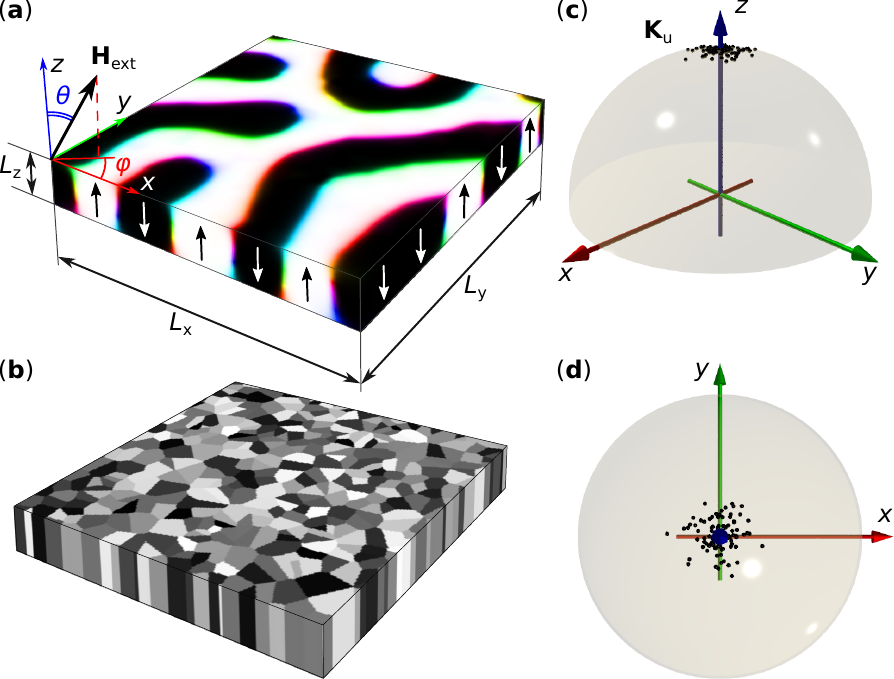}
    \caption{
(a) Simulated domain of square shape with $L_\mathrm{x}=L_\mathrm{y}=464$ nm and thickness $L_\mathrm{z}=54.72$. The external magnetic field orientation is parametrized by angles $\theta$ and $\varphi$. (b) illustrates the underlying grained structure of the film with an average grain size of 25 nm in the $xy$-plane. The grains penetrate through the whole thickness.
(c-d) Representative distribution of uniaxial anisotropy axes in different grains approximated by a random normal distribution.
For details, see the main text and Mumax script in the Supplemental material. 
\label{FigTh1}
}
\end{figure}

\begin{figure*}[ht]
    \includegraphics[width=1.0\linewidth]{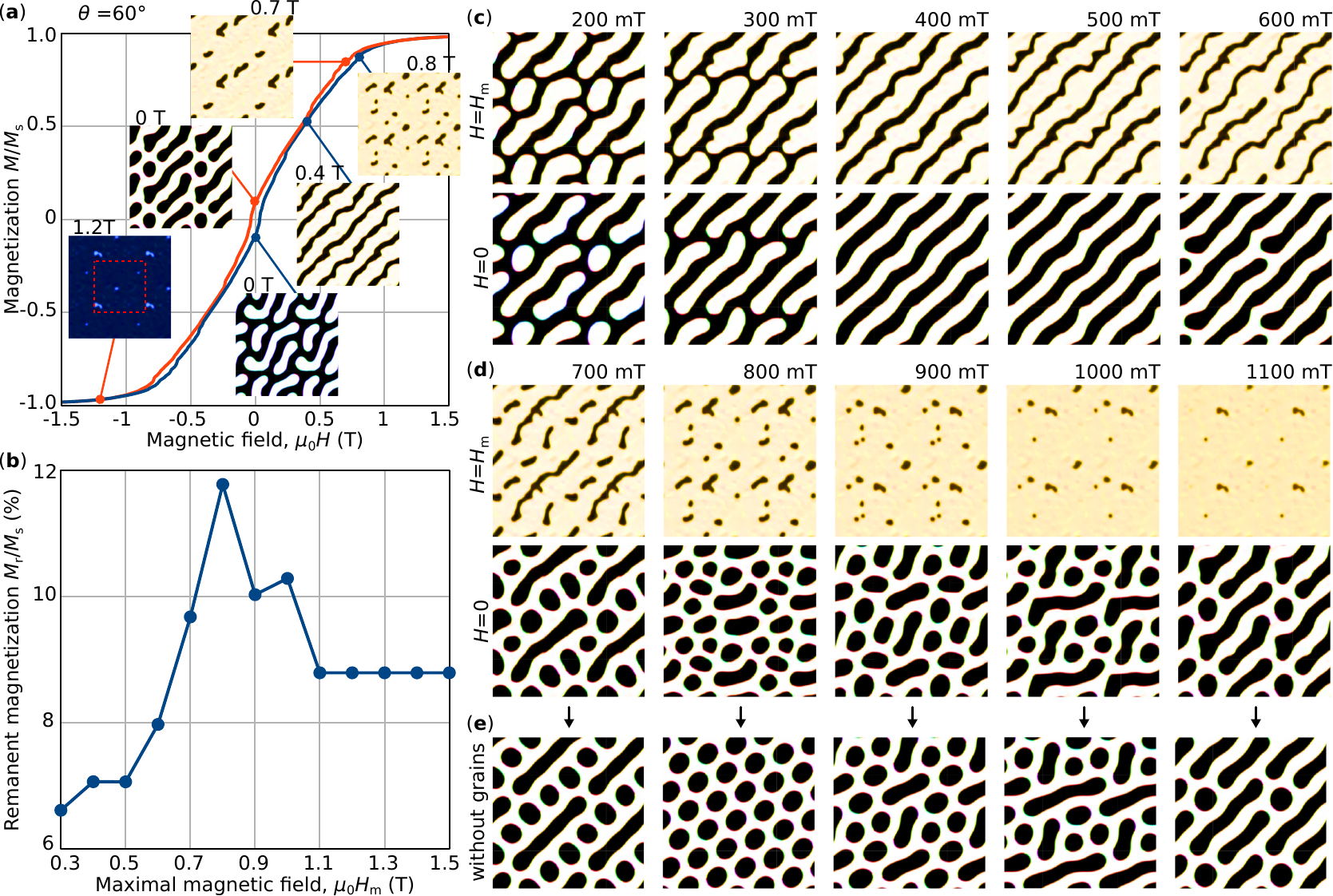}
    \caption{
(a) Magnetization loop obtained in micromagnetic simulations at $\theta=60^\circ$ and $\varphi=45^\circ$. Insets show the snapshots of the magnetization in the center plane of the film.
(b) Theoretically estimated dependence of the remanent magnetization as a function of the maximal magnetic field.
Each pair of images in (c) and (d) has the following meaning: the top image illustrates the magnetization in the middle plane at the maximal magnetic field, while the bottom image shows the state after gradual field reduction to zero. The remanent magnetization of these states is shown in (b).
The images in (e) illustrate the states obtained after full energy minimization with the initial configuration as in the bottom row in (d) and with setting the anisotropy axis in each grain to be parallel to the $z$-axis, which is effectively equivalent to the grains being absent, i.e., the case of an isotropic medium in the $xy$-plane.
\label{FigTh2}
}
\end{figure*}

For micromagnetic simulations, we use a square-shaped area domains in the $xy$-plane with the thickness $L_\mathrm{z}=52.8$\,nm, which is approximately the thickness of the ML with $X=48$.
The size of the sample is chosen to fit the equilibrium period of the stripe domains in the ground state.
We use periodical boundary conditions in the $xy$-plane and parametrize the direction of the applied magnetic field by polar angle $\theta$ and azimuthal angle $\varphi$.
Figure \ref{FigTh1}(a) illustrates the geometry of the simulated domain structure with the maze-like stripe domain state, obtained after energy minimization at random distribution of magnetization in the initial state and $\theta=0^\circ$. 
The obtained stripe domain pattern mimics the maze-like domain texture observed in the experiment, as shown in the top left inset in Fig.~\ref{fig:Fig1}(a).
We use an approximation of continuum anisotropic media, which represents an extension of the earlier used model in Ref.~\cite{Salikhov}. 
Since the coupling between Co layers across Pt layers is weaker than the direct exchange within the Co layer, we assume that the exchange interaction is anisotropic.
In this case the Heisenberg exchange interaction in the micromagnetic energy density functional can be written as
\begin{equation}
w_\mathrm{ex}=A
\sum_i\!\left[
\left( \dfrac{\partial m_i}{\partial x}\right)^{\!2}\!+\!
\left( \dfrac{\partial m_i}{\partial y}\right)^{\!2}\!+
k_\mathrm{iec}\left( \dfrac{\partial m_i}{\partial z}\right)^{\!2}
\right],
\label{FigTh2}
\end{equation}
where $\mathbf{m}=\mathbf{M}/M_\mathrm{s}$ is the magnetization unit vector field and the summation runs over the spatial coordinates $i=x,y,z$. The constant $A$ stands for the exchange coupling in the plane of the film, while the parameter $k_\mathrm{iec}$ defines the weakening of the exchange coupling in the vertical direction due to the Co/Pt multilayering.
In the simulations, which also include demagnetizing fields effect, we use the experimentally estimated values for saturation magnetization, $M_\mathrm{s}=0.775$\,MA/m, uniaxial anisotropy constant, $K_\mathrm{u}=0.6$\,MJ/m$^3$, and
the unitless constant $k_\mathrm{iec}=0.2$, as were estimated earlier in Ref.~\cite{Salikhov} for similar multilayer systems.
In the following, we assume that the film has a granular structure, as illustrated in Fig.~\ref{FigTh1}(b). 
In the first approximation, we exclude the variation of the exchange coupling constant and saturation magnetization between the grains and at the grain boundaries. Thereby, with respect to the exchange
interaction, the media remains continuous.
The grains have an average lateral size of 25 nm and penetrate through the whole film thickness.
This assumption is justified by many experimental studies which confirm column-like shapes of the grains in Co/Pt multilayers~\cite{Ho,Czigany,Pierce,Bohm}.
It is assumed that the anisotropy axis in different grains deviates from the perpendicular direction (Fig.~\ref{FigTh1}(c)-(d)).
The deviation of the polar angle of the vector $\mathbf{K}_\mathrm{u}$ is described by a normal distribution, while the azimuthal angle is uniformly distributed across the interval $[0, 2\pi)$.
Such an assumption of the anisotropy axes is justified by the 
columnar grain growth in polycrystalline MLs, which triggers a wavy surface morphology, known as a correlated roughness in multilayered films, grown by magnetron sputter deposition~\cite{Ho,Czigany,Pierce,Bohm}.
The absolute value of the anisotropy $K_\mathrm{u}$ is fixed in all grains. 
For details of the implementation of the above model in mumax, we refer the reader to the script provided in the Supplemental material~\cite{suppl}.

The representative results of the micromagnetic calculations are presented in Fig.~\ref{FigTh2}. 
We consider the case of a magnetic field tilt angle $\theta=60^\circ$ and $\varphi=45^\circ$. 
First, we simulated a major magnetization loop with maximal field $\pm 1.5$ T. 
We start with the negative saturation at $-1.5$ T, then we gradually increase the field in steps of $0.01$ T and minimize the energy at each step, see the blue curve in Fig.~\ref{FigTh2}(a).
After reaching nearly the saturated state at $+1.5$ T, we gradually decrease the field with the same steps, see the red curve.
We collect snapshots of the magnetization vector field at each field step.
Some representative images of the magnetization in the middle plane of the simulated box are shown in the insets in Fig.~\ref{FigTh2}(a).
Bright and dark contrast indicate positive and negative $z$-component of magnetization, respectively.
We use the standard red-green-blue color code for the in-plane component of magnetization implemented in mumax.
For illustrative purposes, all snapshots of the system depicted in Fig.~\ref{FigTh2}(a), (c)-(e) correspond to
the domain of a larger size,
$2L_\mathrm{x}\times2L_\mathrm{y}$, made by taking into account periodical boundary conditions in the $xy$-plane. The actual size of the simulated domain is indicated for reference by the red dashed square in the inset for $H=-1.2$ T in Fig.~\ref{FigTh2}(a).

As follows from the snapshots in Fig.~\ref{FigTh2}(a), after the saturation at a high magnetic field, the magnetic textures at zero filed always represent a mixture of elongated and short domains.
Note, because of the random distribution of anisotropy axes and as a result of the broken symmetry of the system, the magnetic states at remanence are similar but not identical; compare the two insets for zero field. 
One can also see that the elongated stripe domains are typically aligned with the in-plane projection of the external magnetic field.
At about 0.4 T, the system tends to form a regular stripe domain pattern.
With increasing the field, the regular stripe domains split into isolated domains, see the inset for $0.8$ T.
The density of domains per unit area varies with the field.
Approaching the field of $1.1$ T, most of the domains collapse and only a small fraction of the domains survives above that critical field.
That is illustrated, for instance, by the inset for $-1.2$ T in Fig.~\ref{FigTh2}(a). 
Approaching fields of $\pm1.5$ T, even these domains disappear.

To reproduce the experimental measurements of the remanent magnetization, we perform the following steps.
We start the simulations at different positive magnetic field, $H_\mathrm{m}$, and initial states corresponding to the snapshot of the system at the increasing branch of the magnetization loop (Fig.~\ref{FigTh2}(a)).
The snapshots of these states are depicted in the top row of the images in Fig.~\ref{FigTh2}(c) and (d).
Next, we perform the energy minimization with gradually decreasing field down to zero in steps of 0.01 T.
The examples of the resulting minor magnetization curves are provided in the Supplemental material~\cite{suppl}. 
The remanent magnetization of the system in the projection on the field direction for different  maximal fields $H_\mathrm{m}$ is shown in Fig.~\ref{FigTh2}(b) and the corresponding snapshots of the system at maximal field and at remanence are provided in Figs.~\ref{FigTh2}(c) and (d).
In agreement with the experimental observation, the theoretical dependence of $M_\mathrm{r}(H_\mathrm{m})$ has a distinct maximum at $\sim 0.8$ T.
As follows from the snapshots in Fig.~\ref{FigTh2}(c) and (d), the domain density per simulated domain structure is maximal at this value of $H_\mathrm{m}$ as well as after reduction of the field down to zero.
For $H_\mathrm{m}<0.8$ T, the remanent magnetization and the domain density continuously decay.
On the other hand, with $H_\mathrm{m}$ exceeding the critical value $\sim 1.1$ T, the remanent magnetization converges to a constant value of $\sim 8.8$\% of saturation magnetization $M_\mathrm{s}$, which also agrees well with the experimental observations.

Besides the consistency with in-field observations, the grained structure also explains the broad distribution of domain sizes depicted in Fig.~\ref{fig:Fig4}, at least on a qualitative level.
To illustrate the role of the grained structure on the domain size distribution, we performed the energy minimization of the magnetic texture shown in the bottom row of the images in Fig.~\ref{FigTh2}(d), with setting the uniaxial anisotropy parallel to the $z$-axis in all grains, which effectively corresponds to an absence of the grain structure.
The resulting domain profile is shown in Fig.~\ref{FigTh2}(e).
It is seen that in the absence of the grains, the system tends to have more regular domains of identical shape and size.
The latter becomes most prominent for $H_\mathrm{m}=0.8$ T, where, in the absence of grains, the system converges to a regular closest packed quasi hexagonal lattice of the magnetic bubble domains.
In this case, the bubbles are slightly elongated only because the square shape simulated domain is not perfectly commensurate with the period of the hexagonal bubble lattice.

It is worth noting that at a high magnetic field, all domains are type-II bubbles~\cite{Malozemoff}, see top row of the images in Fig.~\ref{FigTh2}(d).
With decreasing the external field to zero, most of these domains converge to a transient state with a pair of Bloch points and then transit into normal type-I bubbles at negative magnetic fields.
Such topological transition represents an interesting phenomenon that lies out of the scope of the present study and will be discussed elsewhere.

The agreement between theoretical and experimental results allows us to conclude that the formation of a dense bubble domain lattice under tilted external magnetic fields can be effectively explained by the underlying granular structure of magnetic multilayers synthesized by magnetron sputter deposition.
Without the implementation of the grained structure, neither magnetization curve nor magnetic configurations show agreement with the experimental observations. 
For instance, the spontaneous nucleation of domains, in this case, takes place in a  reverse field only.
It is worth noting that the splitting of the stripe domains into individual bubbles in the model without grains also takes place~\cite{Montoya2, Wu2, Chesnel, Yu_12}.
However, the transition fields turn out to be very sensitive to nonphysical parameters of the model, such as mesh density or field step. 
Transitions of that kind are typically ignored as numerical artifacts.
Moreover, the absolute values of the transition fields between stripes and type-II bubbles in the model without grains are much higher than the experimental values. 
On the contrary, even a quite simplified model of the grain structure provides quantitative agreement with experimental observations.

\section{Conclusions}

In conclusion, we demonstrate an approach for the stabilization of magnetic bubble states in metallic ML films with strong PMA. The approach is based on monitoring the remanent magnetization while performing systematic magnetic field protocols at different tilt angles with respect to the films surface. 
We demonstrate that the remanent magnetisation mimics the domain density, showing a distinct peak for the densest bubble state achievable, which corresponds here for the X = 48 sample to a bubble density of 3736 domains/100\,$\mu$m$^2$.
Furthermore, our approach allows the identification of different types of bubble states at ambient conditions within one and the same sample. These states are characterized by different mean bubble sizes and size distribution. These characteristics can be tuned by choosing a suitable magnetic field angle and amplitude.
The micromagnetic modeling provides a quantitative agreement with experimental observations and suggests that the granular structure of the MLs is responsible for the broad distribution of domain sizes observed experimentally.
Our work provides the foundation for further exploration of topological switching of DWs in  dipolar bubbles, which are stabilized in metallic MLs at room temperature and zero magnetic field.

\section{Materials and Methods}

The [(Co(0.44 nm)/Pt(0.7 nm)]$_X$, $X = 48$, 100, 150 MLs films were fabricated at room temperature by DC magnetron sputter deposition at 0.4 Pa Ar atmosphere in an ultrahigh vacuum system ATC 2200 from AJA International Inc. Si substrates with 100 nm thick thermally oxidized (SiO$_2$) layer were used. Prior to the multilayer deposition, a 1.5 nm Ta layer was used for adhesion purposes. A subsequent 20 nm Pt serves as a buffer layer in order to obtain a preferred (111)-texture for the Co/Pt MLs, which supports better growth and larger PMA. The sample was finally capped by a 2 nm Pt layer to avoid surface oxidation. Magnetic measurements were performed using a commercial Microsense EZ7 vibrating sample magnetometer (VSM), equipped with an electromagnet, which delivers up to 1.8 T magnetic field, and with a $\theta=360^\circ$ rotational stage. Magnetic domain imaging was performed using a magnetic force Bruker Dimension Icon microscope. Magnetic images were analyzed using Gwiddion software, which employs a magnetic contrast marking (watershed) algorithm and thus allows an accurate calculation of the domain density and area of a particular magnetization direction. %
All MFM images were recorded at room temperature and zero magnetic field.
The saturation magnetization of all samples was measured using the VSM and calculated to be $M_\mathrm{s}$ = 0.77 ± 0.07\,MA/m. The PMA constant was determined from the area between IP and OOP hysteresis loops to $K_\mathrm{u} \approx 0.6$\,MJ/m$^3$, leading to $Q \approx 1.6$.

\section{Acknowledgments}

The authors are grateful to Thomas Naumann and Jakob Heinze for experimental and technical support.

\end{document}